*Review*

# 3D Deep Learning on Medical Images: A Review

Satya P. Singh [1,2], Lipo Wang [3], Sukrit Gupta [4], Haveesh Goli [4], Parasuraman Padmanabhan [1,2,*] and Balázs Gulyás [1,2,5]

1. Lee Kong Chian School of Medicine, Nanyang Technological University Singapore, 608232, Singapore; satya@ntu.edu.sg (S.P.S.); balazs.gulyas@ntu.edu.sg (B.G.)
2. Cognitive Neuroimaging Centre, Nanyang Technological University Singapore, 636921, Singapore
3. School of Electrical and Electronic Engineering, Nanyang Technological University Singapore, 639798, Singapore; elpwang@ntu.edu.sg
4. School of Computer Science and Engineering, Nanyang Technological University Singapore, 639798, Singapore; SUKRIT001@e.ntu.edu.sg (S.G.); HAVEESH001@e.ntu.edu.sg (H.G.)
5. Department of Clinical Neuroscience, Karolinska Institute, 17176 Stockholm, Sweden
* Correspondence: ppadmanabhan@ntu.edu.sg

**Abstract:** The rapid advancements in machine learning, graphics processing technologies and the availability of medical imaging data have led to a rapid increase in the use of deep learning models in the medical domain. This was exacerbated by the rapid advancements in convolutional neural network (CNN) based architectures, which were adopted by the medical imaging community to assist clinicians in disease diagnosis. Since the grand success of AlexNet in 2012, CNNs have been increasingly used in medical image analysis to improve the efficiency of human clinicians. In recent years, three-dimensional (3D) CNNs have been employed for the analysis of medical images. In this paper, we trace the history of how the 3D CNN was developed from its machine learning roots, we provide a brief mathematical description of 3D CNN and provide the preprocessing steps required for medical images before feeding them to 3D CNNs. We review the significant research in the field of 3D medical imaging analysis using 3D CNNs (and its variants) in different medical areas such as classification, segmentation, detection and localization. We conclude by discussing the challenges associated with the use of 3D CNNs in the medical imaging domain (and the use of deep learning models in general) and possible future trends in the field.

## 1. Introduction

Medical images have varied characteristics depending on the target organ and the suspected diagnosis. Common modalities used for medical imaging include X-ray, computed tomography (CT), diffusion tensor imaging (DTI), positron emission tomography (PET), magnetic resonance imaging (MRI), and functional MRI (fMRI) [1–4]. In the past thirty years, these radiological image acquisition technologies have enormously improved in terms of acquisition time, image quality, resolution [5–8] and have become more affordable. Despite improvements in hardware, all radiological images require subsequent image analysis and diagnosis by trained human radiologists [9]. Besides the significant time and economic costs involved in training radiologists, radiologists also suffer from limitations due to their lack of experience, time and fatigue. This becomes especially significant because of an increasing number of radiological images due to the aging population and more prevalent scanning technologies that put additional stress on radiologists [9–12]. This puts a focus on automated machine learning algorithms that can play a crucial role in assisting clinicians in alleviating their onerous workloads.

Deep learning refers to learning patterns in data samples using neural networks containing multiple interconnected layers of artificial neurons [11]. An artificial neuron by analogy to a biological neuron is something that takes multiple inputs, performs a simple computation and produces an output. This simple computation has the form of a linear function of the inputs followed by an activation function (usually non-linear). Examples of some commonly used non-linear activation functions are the hyperbolic tangent (tanh), sigmoid transformation and the rectified linear unit (ReLU) and their variants [13]. The development of deep learning can be traced back to Walter Pitts and Warren McCulloch (1943). Their work has been followed by significant advancements due to the development of the backpropagation model (1960), convolutional neural networks (CNN) (1979), long short-term memory (LSTM) (1997), ImageNet (2009) and AlexNet (2011) [14]. In 2014, Google presented GoogLeNet (Winner of ILSVRC 2014 challenge) [15], which introduced the concept of inception modules that drastically reduced the computational complexity of CNN. Deep learning is essentially a reincarnation of the artificial neural network where we stack layer upon layer of artificial neurons. Using the outputs of the terminal layers built on the outputs of previous layers, we can start to describe arbitrarily complex patterns. In the CNN [14], network features are generated by convolving kernels in a layer with outputs of the previous layers, such that the first hidden layer kernels perform convolutions on the input images. While the features captured by early hidden layers are generally shapes, curves or edges, deeper hidden layers capture more abstract and complex features.

Historical methods for automated classification of images involves extensive rule-based algorithms or manual feature handcrafting [16–21], which are time-consuming, have poor generalization capacity and require domain knowledge. All this changed with the advent and demonstrated the success of CNNs. CNNs are devoid of any manual feature handcrafting, require little preprocessing and are translation-invariant [22]. In CNNs, low-level image features are extracted by the initial layers of filters and progressively higher features are learnt by successive layers before classification. The commonly seen X-ray is an example of a two-dimensional (2D) medical image. The machine learning of these medical images is no different from CNNs applied to classify natural images in recent years, e.g., the ImageNet Large Scale Visual Recognition Competition [14]. With decreasing computational costs and powerful graphics processing (units (GPUs) available, it has become possible to analyze three-dimensional (3D) medical images, such as CT, DTI, fMRI, Ultrasound and MRI scans [14] using 3D deep learning. These scans give detailed three-dimensional images of human organs and can be used to detect infection, cancers, traumatic injuries and abnormalities in blood vessels and organs. The major drawback in the application of 3D deep learning on medical images is the limited availability of data and high computational cost. Further, there is a problem of the curse of dimensionality. However, with the recent advancements in neural network architectures, data augmentation techniques and high-end GPUs, it is becoming possible to analyze the volumetric medical data using 3D deep learning. Consequently, since 2012, we have seen exponential growth in the applications of 3D deep learning in different medical image modalities. Here, we present a systematic review of the applications of 3D deep learning in medical imaging with possible future directions. To the best of our knowledge, this is the first review paper of 3D deep learning on medical images.

## 2. Materials and Methods

In a very short time, deep learning techniques have become an alternative to many machine learning algorithms that were traditionally used in medical imaging. We explored various terms used in medical imaging literature to understand the trend in using deep learning in medical imaging applications. We searched for 'machine learning + medical' in the title and abstract in PubMed publication database (on 9 July 2020) and across a predictable trend of using more and more similar data in different approaches (Figure 1). We observed a similar trend for the query 'deep learning + medical', albeit with few publications before 2015. However, while searching for the query '3D deep learning + medical' in the title and abstract, we see a different scenario. An exponential increase can be seen for 'deep learning' and '3D deep learning' after 2015 and 2017 onwards, respectively. This signifies that, while there was not much work in the domain a few years ago, there has been an accelerated rise in the number of publications related to deep learning for both 2D and 3D images.

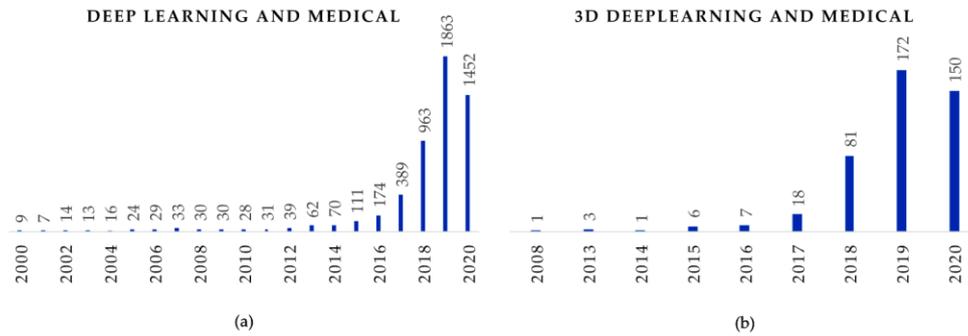

**Figure 1.** Year-wise number of publications in PubMed while searching for 'deep learning + medical' and '3D deep learning + medical' in the title and abstract in PubMed publication database (as at 1st July 2020).

In this systematic review, we searched for the applications of 3D deep learning in medical image segmentation, classification, detection and localization. For the literature search, we chose three database platforms, namely Google Scholar, PubMed and Scopus. The application of 3D CNN effectively came into the picture after the remarkable success of AlexNet in 2012, which was enabled by advanced parallel computing architecture. Between 2015 and 2016, we have seen exponential growth in the literature related to 3D deep learning in medical imaging, and therefore, we limited our search to after 1 January 2012. The first search was performed on 12 September 2019, and the second search on 1 January 2020, while the third search was performed on 1 July 2020. The literature search and selection for the study were done according to the preferred reporting items for systematic seviews and meta-analyses (PRISMA) statement [23]. We searched for title and abstract with different keyword combination of "3D CNN", "medical imaging", "classification", "segmentation", "detection" and "localization" and selected 31,576 records. 11,987 duplicate records were removed. After studying the title and abstract, we further removed 19,380 records. We further excluded 77 records. Finally, we collected 132 papers for our review purpose. The details about the inclusion and exclusion of papers according to the PRISMA statement is depicted in Figure 2.

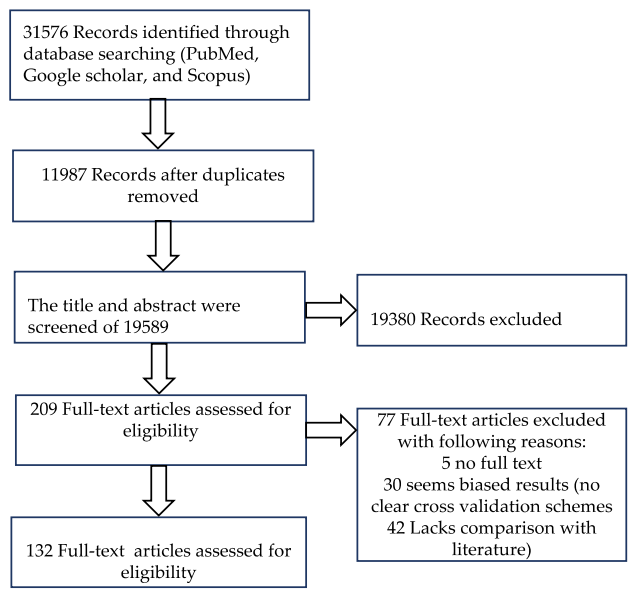

**Figure 2.** Criteria for literature selection for systematic review according to preferred reporting items for systematic reviews and meta-analyses (PRISMA) [23] guidelines.

*2.1. A Typical Architecture of 3D CNN*

A typical architecture of CNN may include four basic components: (1) local receptive field, (2) sharing weights, (3) pooling and (4) fully connected (fc) layers. Deep CNN architecture is constructed by stacking several convolutional layers and pooling layers and one or so fully connected layers at the end of the network [9,24]. While 1D CNN can extract spectral features from the data, 2D CNN can extract spatial features from the input data. However, 3D CNNs can take advantage of both 1D and 2D CNNs by extracting both spectral and spatial features simultaneously from the input volume. These 3D CNN features are very useful in analyzing the volumetric data in medical imaging. The mathematical formulation of 3D CNN is very similar to 2D CNN with an extra dimension added. The basic architecture of 3D CNN is shown in Figure 3. We briefly discuss the mathematical background of 3D CNN.

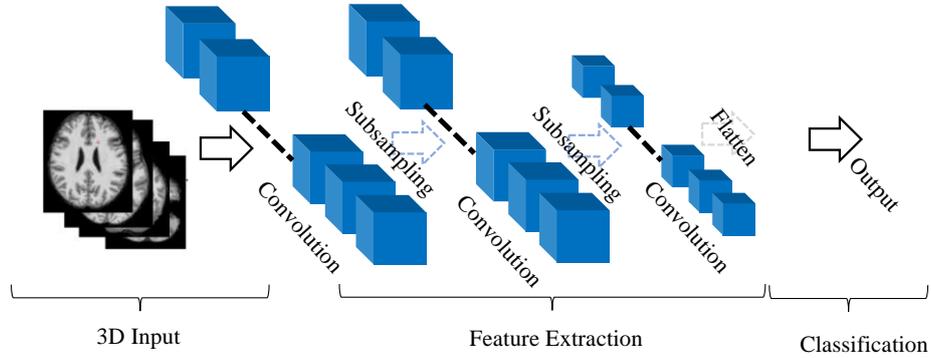

**Figure 3.** Typical architecture of 3D CNN.

*Convolutional Layer*: The basic definition, principle, and working equation of 3D CNN is quite similar to 2D CNN. We only add an extra dimension of depth to the working equation of 2D CNN. Suppose 3D CNN of input $x$ has a dimension of $M \times N \times D$ with $i, j, k$ as iterators. The kernel $\omega$ with dimensions $n1 \times n2 \times n3$ has iterator $a, b, c$. We denote $\ell$ is the $\ell^{th}$, where $\ell = 1$ is the first layer and $\ell = L$ is the last layer. We denote $y^\ell$ and $b^\ell$ as the output and the bias unit the $\ell^{th}$ layer. To compute the nonlinear input $x^\ell_{i,j,k}$ to $(i, j, k)^{th}$ unit in layer $\ell$, we add up the weight contribution from the previous layer as follows:

$$x^\ell_{i,j,k} = \sum_a \sum_b \sum_c \omega_{a,b,c} y^{\ell-1}_{(i+a)(j+b)(k+c)} + b^\ell. \tag{1}$$

The output of the $(i, j)^{th}$ unit in the $'\ell^{th'}$ convolutional layer is given as follows:

$$y^\ell_{i,j,k} = f(x^\ell_{i,j,k}). \tag{2}$$

*Pooling Layer:* Each feature map in the convolutional layer of 3D CNN can be a pooling layer. There are two kinds of pooling. If the pooling layer averages across the group of input voxels, it is called average pooling, while if it obtains a maximum of the input voxels, it is called maximum pooling. The output of the pooling layer will be the input of the next layer. Since a small shift in the input image results in a shift in activation function, the pooling layer also introduces some translational invariance to the 3D CNN. To lower the sampling effect of pooling, we can remove the pooling layer by increasing the number of strides in the preceding CNN layer [25]. This will not result in any significant depreciation of the performance. However, by doing this, we significantly reduce the overlap in the CNN layer that precedes the pooling layer. This is simply equivalent to the pooling operation where only the top-left features are considered.

*Dropout regularization*: Deep neural networks with a large number of parameters are very dominant learning systems. Multiple deep nonlinear hidden layers allow them to learn complex relationships between input and outputs. However, with the limited training data, these complex relationships introduce sampling noise, which appears in training data sets but not in real test datasets even if both are drawn from the same distribution. This scenario leads to overfitting and there have been several strategies [26] to tackle the problem, such as early stopping of the training epochs and weight penalties (L1 and L2 regularizations, soft weight sharing, and pooling). Ensemble models of several CNNs with different configurations on the same dataset are known for their overfitting. However, this leads to extra computational and maintenance cost for training several models. Moreover, training a large network requires large datasets, but the availability of such datasets in the field of medical imaging is very rare. Even if one can train large networks with a versatile setting of parameters, testing these networks is not feasible in a real-time situation due to the nature of medical imaging systems. In the case of ensemble models, a CNN model can also simulate multiple configurations just by probabilistically dropping out edges and nodes. Dropout is a kind of regularization technique to reduce overfitting by temporarily dropping a unit out of the network [27]. This simple idea shows a significant improvement in CNN performance.

*Batch normalization*: The input of each hidden layer dynamically changes during training because the parameters in the previous layer update at each training epoch. If these changes are large, the search for an optimal hyperparameter becomes difficult for the network and may be computationally expensive to reach an optimal value. This problem can be solved by an algorithm called batch normalization, which was proposed by two researchers [28]. Batch normalization allows the use of a higher learning rate and thereby achieves the optimal value in less time. It facilitates the smooth training of deeper network architectures in less time. The normalization of data from a particular batch is about finding the mean and variance of the data points from mini-batch and normalizing them to have a zero mean and unit variance.

In backward pass, the CNN adjusts its weights and parameters according to the output by calculating the error through some loss functions, $e$ (other names are cost function and error function) and backpropagating the error with some rules towards the input. The loss is calculated by taking the partial derivative of $e$ w.r.t., which is the output of each neuron in that layer, such as $\partial e/y^{\ell}_{i,j,k}$ for the output, $y^{\ell}_{i,j,k}$ of $(i,j,k)^{th}$ unit in layer $\ell$. The cFhain rule allows us to write and add up the contribution of each variable as follows:

$$\frac{\partial e}{\partial x^{\ell}_{i,j,k}} = \frac{\partial e}{\partial y^{\ell}_{i,j,k}} \frac{\partial f(y^{\ell}_{i,j,k})}{\partial x^{\ell}_{i,j,k}} = \frac{\partial e}{\partial y^{\ell}_{i,j,k}} f'(x^{\ell}_{i,j,k})\#. \tag{3}$$

Weights in the previous convolutional layer can be updated by backpropagating the error to the previous layer according to the following equation:

$$\frac{\partial e}{\partial y^{\ell-1}_{i,j,k}} = \sum_{a=0}^{n1-1}\sum_{b=0}^{n2-1}\sum_{c=0}^{n3-1} \frac{\partial e}{\partial x^{\ell}_{(i-a),(j-b),(k-c)}} \frac{\partial x^{\ell}_{(i-a),(j-b),(j-b)}}{\partial y^{\ell-1}_{i,j,k}}. \tag{4}$$

$$= \sum_{a=0}^{n1-1}\sum_{a=0}^{n2-1}\sum_{b=0}^{n3-1} \frac{\partial e}{\partial x^{\ell}_{(i-a),(j-b),(k-c)}} \omega_{a,b,c}. \tag{5}$$

Equation (5) allows us to calculate the error for the previous layer. Further, the above eq. makes sense for those points which are n times away from each side of the input data. This situation can be avoided by simply padding with zeros to the end of each side of the input volume.

*2.2. Breakthroughs in CNN Architectural Advances*

Several different versions of CNN have been proposed in the literature to improve model performance. In 2011, Krizhevsky et al. [14] presented a deep CNN architecture. A systematic architecture of AlexNet is shown in Figure 4. AlexNet has five convolutional layers and three fully connected layers (the last FC layer was the SoftMax layer). The network was trained on 1.2 million images with 60 million parameters. To tackle these large parameters, AlexNet was trained on a multi-GPU (2-GPUs, 3GB GTX-580) environment by systematically distributing the neurons on both the GPUs. Data augmentation and dropouts were used to avoid overfitting. Data augmentation was done in two ways: (1) image translations and horizontal reflections, and (2) changing the intensity of RGB channels. The AlexNet architecture has won ILSVRC-2012 (ImageNet Large Scale Visual Recognition Competition-2012) with a large margin. The difference between the top-five test errors of AlexNet (15.3%) and the second prize winner (26.2%) was around 10%.

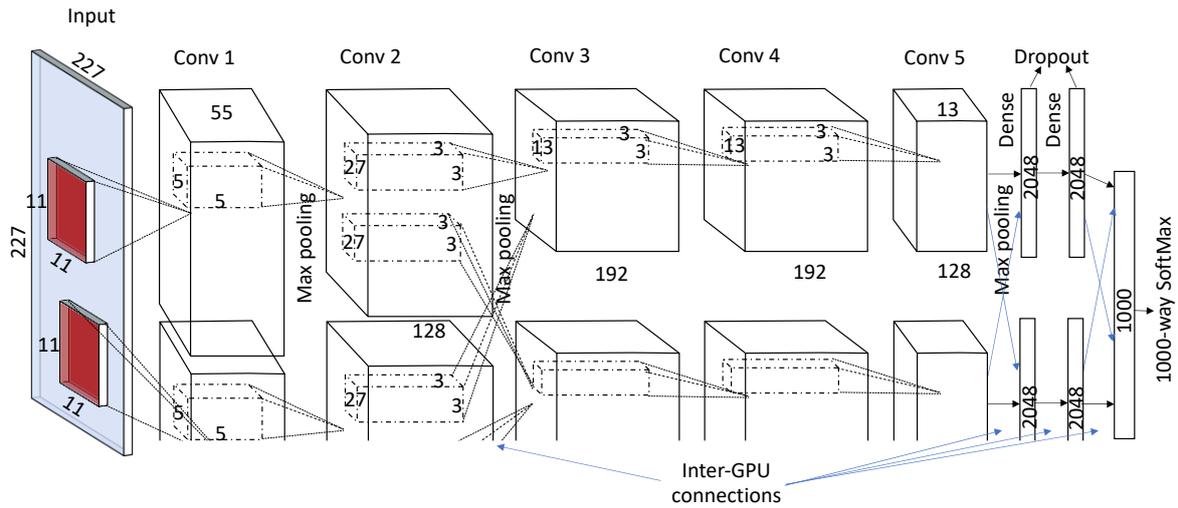

**Figure 4.** A typical architecture of AlexNet [14].

In 2014, Simonyan and Zisserman [29] presented a more profound deep network architecture called VGGNet (Visual Geometry Group Network) (16 layers and 19 layers) for the ImageNet Challenge 2014 and secured the first position for the localization task and the second position for the classification task. VGGNet uses 3 × 3 filters for convolutional layers and three consecutive fully connected layers 4096, 4096, and 1000 in size, respectively. The design of VGGNet is quite similar to AlexNet. Adding consecutive layers to the network increases the number of parameters that cause networks to suffer from errors and overfitting. In supervising learning, the deeper network requires large data for training and despite the use of data augmentation techniques, it may happen that the data is not sufficient. Further, annotating such a large amount of data can be quite expensive. Furthermore, because a linear increase in the filter emerges as a result of a quadratic increase in computational burden, deeper networks lead us to a computational explosion. In deeper layers, weights can be near zero and emerge as a waste of computational resources. Fast forwarding from 2012, in 2014, the designers at Google presented the concept of "Inception" based on the Hebbian principle and the intuition of multiple-scale processing, and the network was called GoogLeNet (also known as Inception-V1) [15]. The intuition behind the inception module (version V1) (Figure 5a) is that the optimal neural network topology can be built by clustering neurons to the correlation statistics in the input images. The authors analyzed the correlation statistics in the activations in the previous layer and clustered the neurons with highly correlated outputs for the next layer. In the images, the correlation tends to be local, and therefore, performing convolutions over the local patches can cluster the neurons. In lower layers, there exists a high correlation between local pixels in a surrounding patch. These pixels can be covered by a small 1 × 1 convolution. Besides, the correlation between a smaller number of spatially spread-out clusters can be quantified by 3 × 3 and 5 × 5 convolutions. In order take effect of 1 × 1, 3 × 3 and 5 × 5, the authors stacked them as a single output vector for the next layer (Figure 5a). In addition, to take the advantages of maximum (max)-pooling, they also concatenated a pooling layer in parallel (3 × 3 max pooling branch in Figure

5a). The model size of GoogLeNet was 12 times less than AlexNet's and required relatively net-lower memory and power than AlexNet. In addition, GoogLeNet's computational cost was less than 2× that of AlexNet, while AlexNet was a network of eight layers and GoogLeNet was a network of 22 layers. Due to its high accuracy, fewer parameters and low power consumption, GoogLeNet was more suitable for mobile platforms. GoogLeNet has secured the first position in the ImageNet Large-Scale Visual Recognition Challenge 2014 (ILSVRC14). The concept of the inception module in GoogLeNet alleviates the problem of vanishing gradients and allows us to move deeper into the network. In the event of increasing computational complexity, the authors suggest reducing network dimensions by introducing Inception-V2 and Inception-V3. [30]. In inception-V2, inexpensive 1 × 1 convolution convolutions were inserted before the expensive 3 × 3 and 5 × 5 convolutions. In addition, in this module, the 1 × 1 convolution used ReLU activation. It was also suggested that the incorporation of inception layers would benefit if inserted in higher-order layers. Since the architecture of the GoogLeNet was rather deep, the designers were also concerned about the problem of gradient vanishing in the hidden layers during back-propagation. Since the intermediate layers in the network had a higher representation of the input image, they had more discriminative power. In Reference [31], the authors presented the concept of Inception-V4 by appending the auxiliary classifiers to the intermediate layers. Since the FC connected layers were generally prone to error, in order to avoid overfitting, the authors replaced the FC layers with average pooling layers.

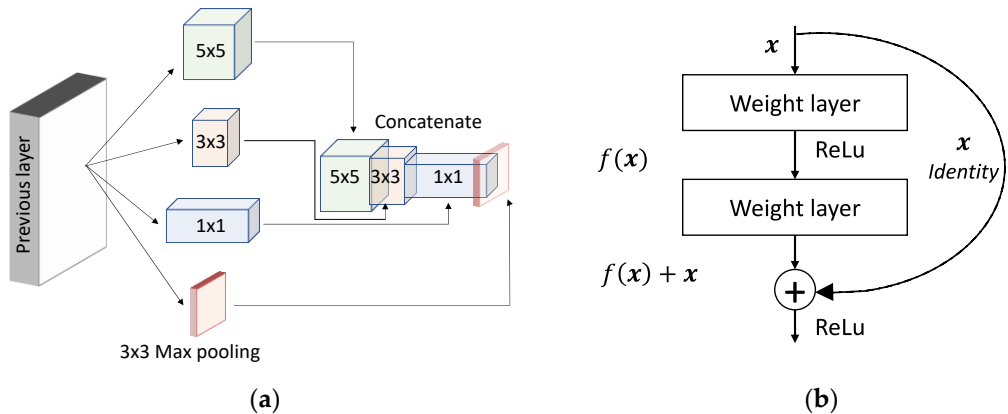

**Figure 5.** (**a**) The intuition behind the inception (V1) module in GoogLeNet. Screening local clusters with 1 × 1 convolutional operations, screening spread-out clusters with 3 × 3, screening even more spread-out clusters with 5 × 5 convolutional operations, and finally conceiving the inception module by concatenating (**b**) residual building block in ResNet.

Very deep architectural neural networks are often difficult to train due to the problem of vanishing and exploding gradients. Considering a shallow CNN and its deeper counterpart with more layers, a deeper model is theoretically just needed to copy the output from the shallow model with identity mappings. Therefore, the constructed solution suggests that the deeper model must not produce higher errors than the shallow model. However, the identity functions are not easy to learn, and therefore, in Reference [32], He et al. presented ResNet by reformulating the layers as residual learning functions. ResNet uses skip connections, which allows us to take the activation from one layer and suddenly feed it to another layer. Original ResNet uses batch normalization after each convolutional layer and before the activation. Using these skip connections, we can train very deep network architectures, ranging from 52 to 10,000 layers. ResNet is one of the most popular deep learning architectures in the literature. Figure 5b shows the basic concept of residual connection, which is the building block of ResNet. The authors trained 152 layers deep ResNet (11.3 billion FLOPs) on ImageNet against VGGNet-19 (15.3/19.3 billion FLOPs) and obtained state-of-the-art results. This was about eight times deeper than VGGNet, but in terms of the floating-point operation measurement, it has less computation. With a 3.57% test error on the ImageNet dataset, ResNet secured first place in ILSVRC-2015. Inspired by the power of the inception module from GoogLeNet and residual connections from ResNet [32], the research team from Google further presented

Inception-ResNet [31]. They showed that the residual connections accelerate training in the inception network and thereby improved the performance of the Inception network. Continuous concatenation of convolutional layers on the top of activations will make the training worse. Ronneberger et al. [33] proposed U-Net, which was the winner of the ISBI cell tracking challenge 2015 for biomedical image segmentation tasks. U-Net can be trained with a relatively small number of samples and achieves high accuracy in a short time. U-Net involves the symmetrical expansion and contraction paths with skip connections. In 2016, Cicek et al. [34] presented a modified version of the original U-Net i.e., 3D U-Net for volumetric segmentation from sparse notation.

**3. 3D Medical Imaging Pre-Processing**

Preprocessing of the image dataset before feeding the CNN or other classifiers is important for all types of imaging modalities. Several preprocessing steps are recommended for the medical images before they are fed as input to the deep neural network model, such as (1) artifact removal, (2) normalization, (3) slice timing correction (STC), (4) image registration and (5) bias field correction. While all the steps, (1) to (5), help in getting reliable results, STC and image registration are very important in the case of 3D medical images (especially MR and CT images). Artifact removal and normalization are the most performed preprocessing steps across the modalities. We briefly discuss pre-processing steps above.

The first part of any preprocessing pipeline is the removal of artifacts. For example, we may be interested in removing skulls in brain CT scans before feeding to 3D CNN. Removal of extracerebral tissues is highly recommended before analyzing the T1 or T2 weighted MRI, and DTI modalities for brain images. fMRI data often contains transient spike artifacts or a slowed over drift time. Thus, the principal component analysis technique can be used to look at these spike related artifacts [3,35,36]. Before feeding the data for preprocessing to an automated pipeline, a manual check is also advisable. For example, if the input T1 anatomical data is large, the FSL's BET command will not perform proper brain region extraction, and if we use images with artifacts for the popular fMRI preprocessing tool fMRIprep [37], it fails as well. Therefore, to remove these extra neck tissues, we should perform other necessary preprocessing steps.

The brain and other body parts for the imaging of every person can vary in shape and size. Hence, it is advisable to normalize brain scans before further processing. [4,38–41]. Due to the characteristics of imaging modalities, the same scanning device can essentially have different intensities even in the same patient's medical images. Since scanning of patients may be performed in different light conditions, intensity normalization also plays an important role in the performance of 3D CNN. Besides, for a typical CNN, each input channel (i.e., sequence) is normalized to have a zero mean and unit variance within the training set. Parameter normalization within the CNN also affects the CNN performance.

To create a volumetric representation of the brain, we often sample several slices in the brain for each repetition time (TR). However, each slice is typically sampled at slightly different time points as we acquire them sequentially [42,43]. Hence, even though the 3D brain volume should be scanned instantaneously, in practical terms, there is always some delay in sampling the first and the last slice. This is a key problem that needs to be considered and accounted before performing any further tasks like classification or segmentation. In this regard, STC is frequently employed for adjusting the temporal misalignment and is widely utilized by a range of software such as SPM and FSL [44]. Several types of techniques have been proposed based on data interpolation methods for STC, including cubic spline, linear and CNC interpolation [45]. In general, the STC methods based on interpolation techniques can be grouped as scene-based and object-based. In the scene-based approach, the interpolated pixel intensity is revealed by the pixel intensity of a slice. While the interpolation techniques are sub-standard, they are relatively simple and easy to implement. However, object-based methods have much better accuracy and are reliable, but they are computationally expensive. Subsequently, cubic spline and other polynomials were also found in medical image interpolation. Essentially, all these strategies perform strength averaging of the neighboring pixels without forming any feature deformation. Therefore, the resultant in-between pixels have negative blurring effects within the object boundary. Cubic interpolation is the standard technique selected in BrainVoyager [46] software.

Medical imaging is becoming increasingly multimodal, whereby images of the same patient from different modalities are acquired to provide information about different organ features. Additionally, situations also arise where multiple images of the same patient and location are acquired with different orientations. It was necessary to match the images by visual comparison in this case [47]. This alignment or registration of the images to a standard template can also be automated, which helps to locate repetitive locations of abnormalities. The image alignment not only makes it easier to manually analyze images and locate lesions or other abnormalities, but also makes it easier to train a 3D CNN on these images [48–50].

MRI images are corrupted by a low-frequency and smooth bias field signal produced by MRI scanners, thereby affecting pixel intensities to fluctuate [51,52]. The bias field usually appears due to improper image acquisition from the scanner, and influences machine learning algorithms that perform classification and segmentation using pixel intensities. It is, therefore, important to either remove the bias field artifacts from sample images or incorporate this artifact into the model before training on these images.

## 4. Applications in 3D Medical Imaging

*4.1. Segmentation*

For several years, machine learning and artificial intelligence algorithms have been facilitating radiologists in the segmentation of medical images, such as breast cancer mammograms, brain tumors, brain lesions, skull stripping, etc. Segmentation not only helps to focus on specific regions in the medical image, but also helps expert radiologists in quantitative assessment, and planning further treatment. Several researchers have contributed to the use of 3D CNN in medical image segmentation. Here, we focus on the important related works of medical image segmentation using 3D CNN.

Lesion segmentation is probably the most challenging task in medical imaging because lesions are rather small in most of the cases. Further, there are considerable variations in their sizes across different scans that can cause imbalances in training samples. In this regard, Deep Medic [53] is a popular work, which also won the ISLES 2015 competition. In DeepMedic, a 3D CNN architecture has been introduced for automatic brain lesion segmentation, which gives a state-of-the-art performance on 3D volumetric brain scans. The multiresolution approach has been utilized to include local as well as the spatial contextual information. The network gives a 3D map of where the network believes the lesions are located. DeepMedic was implemented on datasets where patients suffered from traumatic brain injuries due to accidents and were also shown to work well for classification and detection problems in head images to detect brain tumors. This work was carried forward by Kamnitsas et al. [54] during the brain tumor segmentation (BRATS) 2016 challenge where the authors took advantage of residual connections in 3D CNN (Figure 6). The results were impressive and were in the top 20 teams with median Dice scores of 0.898 (whole tumor, WT), 0.75 (tumor core, TC) and 0.72 (enhancing core, EC). Following DeepMedic, Casamitjana et al. [55] proposed a 3D CNN to process the entire 3D volume in a single pass to make predictions.

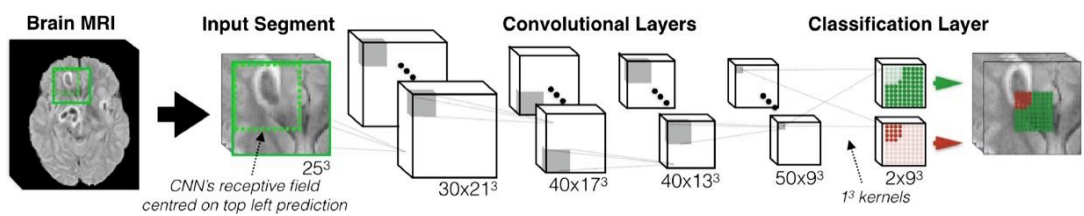

**Figure 6.** The baseline architecture of 3D convolution neural network (CNN) for lesion segmentation. The figure is slightly modified from [54].

Besides constraints in acquiring enough training samples, class imbalance also pervades in the medical imaging domain, whereby samples of the diseased patients are hard to come by. This issue is further exacerbated in problems related to the tumor or lesion segmentation because the sizes of tumors or lesions are usually small when compared to the whole scan volume. In this context, Zhou et al. [56] proposed 3D CNN (3D variant of FusionNet) for brain tumor segmentation on the BRATS 2018 challenge. The authors split the multiclass tumor segmentation problem into three separate segmentation tasks for the deep 3D CNN model, i.e., (i) coarse segmentation for whole tumor, (ii) refined segmentation for Wavelet transform (WT) and intraclass tumor, and (iii) precise segmentation for a brain tumor. Their model has ranked first for the BRATS 2015 dataset and third (among 64 teams) on the BRATS 2017 validation dataset. Ronneberger et al. proposed the U-Net architecture for the segmentation of 2D biomedical images [33] They made use of up-sampling layers, which in turn enabled the architecture for segmentation besides classification. However, the original U-Net was not too deep as there was a single pooling layer after the convolution layer. Further, this only analyzed 2D images and did not fully exploit the spatial and texture information that can be obtained from the 3D volumes. To solve these issues, Chen et al. [57] proposed a separable 3D U-Net for brain tumor segmentation. On BRATS 2018 challenge dataset, they achieved dice scores of 0.749 (EC), 0.893 (WT) and 0.830 (TC). Kayalibay et al. [58] presented a modified 3D U-Net architecture for brain tumor segmentation where they introduce some nonlinearity in the traditional U-Net architecture by inserting residual blocks during up-sampling, thus facilitating the gradients to flow easily. The proposed architecture also intrinsically handles the class imbalance problem that arises due to the use of the Jaccard loss function. However, the proposed architecture was computationally expensive owing to the large size of the receptive field used. Isensee et al. [59] proposed a 3D U-Net architecture, which consists of a perspective collection pathway for brain tumor segmentation. The strategy encodes progressively abstract interpretations of the input as we move deeper and adds a localization pathway that recombines these interpretations with features for lower layers. By hypothesizing that semantic features are easy to learn and process, Peng et al. [60] presented a multi-scale 3D U-Net for brain tumor segmentation. Their model consists of several 3D U-Net blocks for capturing long-distance spatial resolutions. The upsampling was done at different resolutions to capture meaningful features. On the BRATS 2015 challenge dataset, they achieved 0.893 (WT), 0.830 (TC) and 0.742 (EC). Some important developments in 3D CNN for brain tumor/lesion segmentation applications on BRATS challenges are summarized in Table 1.

While brain tumor or lesion segmentation is used to detect glioblastoma, brain stroke or traumatic brain injuries, multiple deep learning solutions are being proposed for the segmentation of brain lobes or deep brain structures. Milletari et al. [61] combined a Hough voting approach with 2D, 2.5D and 3D CNN to segment volumetric data of MRI scans. However, these networks still suffer from the class imbalance problem. In Reference [62], a 3D CNN was implemented for subcortical brain structure segmentation in MRI and this study was based on the effect of the size of the kernels in a network. In Reference [34], the authors applied 3D U-Net for dense volume segmentation. However, this network was not entirely in 3D because it used 2D annotated slices for training. Sato et al. [63] proposed 3D deep network for the segmentation of the head CT volume.

Liver cancer is one of the major causes of cancer deaths worldwide. Therefore, reliable and automated liver tumor segmentation techniques are needed to assist radiologists and doctors in hepatocellular carcinoma identification and management. Duo et al. [64] presented a fully connected 3D CNN for liver segmentation from 3D CT scans. The same network has also been tested on the whole heart and great vessel segmentation. Further, 3D U-Net has been applied in liver segmentation problems [65]. In Reference [66], 3D ResNet has been used for liver segmentation using the coarse-to-fine approach. Some other similar approaches for segmentation of the liver can be found in References [43,67–69]. In this sequence, another work, based on the 2D DenseUnet and hierarchical diagnosis approach (H-DensNet) for the segmentation of liver lesions, has been presented in Reference [70]. This network secured the first position in the LiTS 2017 leaderboard. The network has been tested on the 3D IRCADs database and achieved state-of-the-art outcomes, outperforming the other very well-established liver segmentation approaches. They have achieved dice scores of 0.982 and 0.93.7 for liver and tumor segmentation, respectively.

Table 1. 3D CNN for brain tumor/lesion segmentation on brain tumor segmentation (BRAST) challenges.

| Ref. | Methods | Data | Task | Performance Evaluation |
|---|---|---|---|---|
| Zhou et al. [56] | A 3D variant of FusionNet (One-pass Multi-task Network (OM-Net)) | BRATS 2018 | brain tumor segmentation | 0.916 (WT), 0.827 (TC), 0.807(EC) |
| Chen et al. [57] | Separable 3D U-Net | BRATS 2018 | --do-- | 0.893(WT), 0.830(TC), 0.742(EC) |
| Peng et al. [60] | Multi-Scale 3D U-Nets | BRATS 2015 | --do-- | 0.850(WT), 0.720(TC), 0.610(EC) |
| Kayalıbay et al. [58] | 3D U-Nets | BRATS 2015 | --do-- | 0.850 (WT), 0.872(TC), 0.610(EC) |
| Kamnitsas et al. [54] | 11 layers deep 3D CNN | BRATS 2015 and ISLES 2015 | --do-- | 0.898 (WT), 0.750 (TC), 0.720(EC) |
| Kamnitsas et al. 2016 [53] | 3D CNN in which features extracted by 2D CNNs | BRATS 2017 | --do-- | 0.918 (WT), 0.883(TC), 0.854 (EC) |
| Casamitjana et al. [55] | 3D U-Net followed by fully connected 3D CRF | BRATS 2015 | --do-- | 0.917(WT), 0,836(TC), 0.768(EC) |
| Isensee et al. [59] | 3D U-Nets | BRATS 2017 | --do-- | 0.850(WT), 0.740(TC), 0.640(EC) |

3D CNNs are also being used in the segmentation of knee structures. In Reference [71], Ambellan et al. proposed a technique with 3D statistical shape models along with 2D to accomplish an effective and precise segmentation of knee structures. In Reference [72], the authors suggested a 3D CNN to segment cervical tumors on 3D PET images. Their architecture uses spatial information for segmentation purposes. The authors claimed highly precise results for segmenting cervical tumors on the 3D PET. In Reference [73], the authors proposed 3D convolution kernels for learning filter coefficients and spatial filter offsets simultaneously for 3D CT multi-organ segmentation work. The outcomes were compared to U-Net architectures and the authors claim that their architecture requires less trainable parameters and storage to obtain a high quality. In Reference [74], Chen et al. proposed 3D FC deep CNN (3D UNet) for the segmentation of six thoracic and abdominal organs (liver, spleen, and left/right kidneys and lungs) from dual-energy computed tomographic (DECT) images.

*4.2. Classification*

Classification of diseases using deep learning technologies on medical images has gained a lot of traction in the last few years. For neuroimaging, the major focus of 3D deep learning has been on detecting diseases from anatomical images. Several studies have focused on detecting dementia and its variants from different imaging modalities, including functional MRI and DTI. Alzheimer's Disease (AD) is the most common form of dementia, usually linked to the pathological amyloid depositions, structural-atrophy and metabolic variations in the chemistry of the brain. The timely diagnosis of AD plays an important role to intercept the progression of the disease.

Yang et al. [39] visualized the 3D CNN, trained to classify AD in terms of AD features, which can be a very good step in understanding the behavior of each layer of 3D CNN. They proposed three types of visual inspection approaches: (1) sensitivity analysis, (2) 3D class activation mapping and (3) 3D weighted gradient weighted mapping. The authors explained how visual inspection can improve accuracy and aid in deciding the 3D CNN architecture. In their work, some well-known baseline 2D deep architectures, such as VGGNet and ResNet, were converted to their 3D counterparts, and the classification of AD was performed using MRI data from the Alzheimer's Disease Neuroimaging Initiative (ADNI). In Reference [75], the authors trained an auto-encoder to derive an embedding from the input features of 3D patches. These features were extracted from the preprocessed MRI scans downloaded from the ADNI dataset. Their work demonstrated an improvement in results in comparison to the 2D approaches available in the literature. In Reference [76], the authors stacked recurrent neural network (long short-term memory) layers on 3D CNN layers for AD classification tasks using PET and MRI data. The 3D fully connected CNN layers obtained deep feature representations and the LSTM was applied on these features to improve the performance. In Reference [77], a deep 3D CNN was researched on a sizeable dataset for the classification of AD. Gao et al. [78] showed 87.7% accuracy in the classification of AD, lesion and normal aging by implementing a seven-layer deep 3D CNN on 285 volumetric CT head scans from Navy General hospital, China. In this study, the authors also compared their results from 3D CNN with hand-crafted features of 3D scale-invariant Fourier transform (SIFT) and showed that the proposed 3D CNN approach gives around four percent higher classification accuracy.

**Table 2.** 3D CNN for classification tasks in medical imaging.

| Ref. | Task | Model | Data | Performance Measures |
|---|---|---|---|---|
| Yang et al. [39] | AD classification | 3D VggNet, 3D Resnet | MRI scans from ADNI dataset (47 AD, 56 NC) | 86.3% AUC using 3D VggNet and 85.4% AUC using 3D ResNet |
| Kruthika et al. [75] | --do-- | 3D capsule network, 3D CNN | MRI scans from ADNI dataset (345 AD, NC, 605, and 991MCI) | Acc. for AD/MCI/NC 89.1% |
| Feng et al. [76] | --do-- | 3D CNN + LSTM | PET + MRI scans from ADNI dataset (93 AD, 100 NC) | Acc. 65.5% (sMCI/NC), 86.4% (pMCI/NC), and 94.8 % (AD/NC) |
| Wegmayr et al. [77] | --do-- | 3D CNN | ADNI and AIBL data sets, 20000 T1 scans | Acc. 72% (MCI/AD), 86 % (AD/NC), and 67 % (MCI/NC) |
| Oh et al. [84] | --do-- | 3D CNN +transfer learning | MRI scans from the ADNI dataset (AD 198, NC 230, pMCI 166, and sMCI 101) at baseline. | 74% (pMCI/sMCI), 86% (AD/NC), 77% (pMCI/NC) |
| Parmar et al. [10] | --do-- | 3D CNN | fMRI scans from ADNI dataset (30 AD, 30 NC) | Classification acc. 94.85 % (AD/NC) |
| Nie et al. [79] | Brain tumor | 3D CNN with learning supervised features | Private, 69 patient (T1 MRI, fMRI, and DTI) | Classification acc. 89.85 % |
| Amidi et al. [85] | Protein shape | 2-layer 3D CNN | 63,558 enzymes from PDB datasets | Classification acc. 78% |
| Zhou et al. [80] | Breast cancer | Weakly supervised 3D CNN | Private, 1537 female patient | Classification acc. 78% 83.7% |

Besides detecting AD using head MRI (or other modalities), multiple studies have been performed to detect diseases from varied organs in the body. Nie et al. [79] took advantage of the 3D aspect of MRI by training a 3D CNN to evaluate the survival in patients going through high-grade gliomas. Zhou et al. [80] proposed a weakly-supervised 3D CNN for breast cancer detection. However, there are several limitations of the study: (1) the data was selective in nature, (2) the proposed architecture was only able to detect the tumor with high probability and (3) only structural features were used for the experiments. Jnawali et al. [41] demonstrated the performance of 3D CNN in the classification of CT brain hemorrhage scans. The authors constructed three versions of the 3D architectures based on CNNs. Two of these architectures are 3D versions of the VggNet and GoogLeNet. This unique research was done on a large private dataset and about 87.8% accuracy was demonstrated. In Reference [9], Ker et al. developed a three-layer shallow 3D CNN for brain hemorrhage classification. The proposed network was giving state-of-the-art results with small training time when compared to 3D VGGNet and 3D GoogLeNet. Ha et al. [81] modified 2D U-Net into 3D CNN to quantify the breast MRI fibro-glandular tissue (FGT) and background parenchymal enhancement (BPE). In Reference [58], Nie et al. proposed a multi-channel structure of 3D CNN for survival time prediction of Glioblastoma patients using multi-modal head images (T1 weighted MRI and diffusion tensor imaging, DTI). Recently, in Reference [82], the author presented a hybrid model for the classification and prediction of lymph node metastasis (LNM) in head and neck cancer. They combined the outputs of MaO-radiomics and 3D CNN architecture by using an evidential reasoning (ER) fusion strategy. In Reference [83], the authors presented a 3D CNN for predicting the maximum standardized uptake value of lymph nodes in patients suffering from cancer using CT images from a PET/CT examination. We summarized some important developments in 3D deep learning models for classification tasks in medical imaging in Table 2.

*4.3. Detection and Localization*

Cerebral Microbleeds (CMBs) are small foci of chronic hemorrhages that can occur in the normal brains due to structural abnormalities of small blood vessels in the brain. Due to the differential properties of blood, MRI can detect CMBs. However, detecting cerebral micro-hemorrhages in brain tissue is a difficult and time-consuming task for radiologists, while recent studies employed 3D deep architectures to detect CMBs. Dou et al. [86] proposed a two-stage fully connected 3D CNN architecture to detect CMBs from the dataset of MRI susceptibility-weighted images (SWI). The network reduced many false-positive candidates. For training purposes, multiple 3D cubes were extracted from the preprocessed dataset. This study also examined the effect of the size of 3D patches on network performance. The study also focuses on the higher performance of 3D architectures in the detection of CMBs in comparison to their 2D architectures, such as Random Forest and 2D-CNN-SVM. Dou et al. further employed a fully 3D CNN to detect microscopic areas of a brain hemorrhage on MRI brain scans. This method had a sensitivity of 93% and outperformed prior methods of detection. Standvoss et al. [87] detected CMBs in traumatic brain injury. In their study, the authors prepared three types of 3D architectures with varying depths, i.e., three, five and eight layers. These models were quite simple and straight forward, with an overall best accuracy of 87%. The drawback of these studies was that they utilized a small dataset for training the network. In Reference [88], the author presented a 3D CNN to forecast the route and radius of an artery at any given point in a cardiac CT angiography image, which depends on the local image patch. This approach can precisely and effectively predict the path and the radius of coronary arteries through the details extracted from the image files.

Lung cancer is also the foremost cause of death worldwide. Nonetheless, the survival rate would be increased if we could detect lung cancer at an early stage. Subsequently, the past decade has seen considerable research into the detection, classification and localization of lung nodules using 3D deep learning approaches. In Reference [89], Anirudh et al. first proposed a 3D CNN for lung nodule detection using weakly labeled data. In 3D medical imaging, data labeling is quite complex and time-consuming when compared to 2D image modalities. The authors used a single-pixel point to unveil the data and used this single point information to grow the region using the thresholding and filtering of super-pixels. This process was performed on 2D slices and these slices were combined using 3D Gaussian filtering. Using the proposed 3D CNN, the authors showed an 0.80 sensitivity with 10 false positives per scan. However, the architecture of 3D CNN was not very deep in this work. Furthermore, the data were very small (70 scans), and therefore, the results may be biased. Dou et al. [90] exploited 3D CNN with multilevel contextual information for the false-positive reduction in pulmonary nodules in volumetric CT scans. The authors used 887 CT scans from a publicly available LIDC-IDRI dataset (LUNA16 challenge). Huang et al. [91] exploited

3D CNN to detect lung nodules in low-dose CT chest scans. The positive and negative cubes were extracted from CT data using a priori knowledge about the data and confounding the anatomical structure. The proposed design effectively reduced the complexity and showed a significant improvement in performance. Compared to the baseline approach, their approach showed 90% sensitivity, while a reduction in false positives from 35 to 5. Gruetzemacher et al. [12] used 3D UNet with residual blocks for detecting pulmonary nodules in CT scans from the LIDC-IDRI dataset. The authors used two 3D CNN models, one for each essential task, i.e., candidate generation and false-positive reduction. The model was experimented and evaluated with 888 CT scans. On the test data, an overall 89.3% detection and 1.79 false-positive rate was obtained. To tackle large variations in the size of the nodules, Gu et al. [92] proposed multi-scale prediction with a fusion scheme for 3D CNN (Figure 7). This work was also a part of the LUNA16 challenge and achieved 92.9% sensitivity with four false positives per scan.

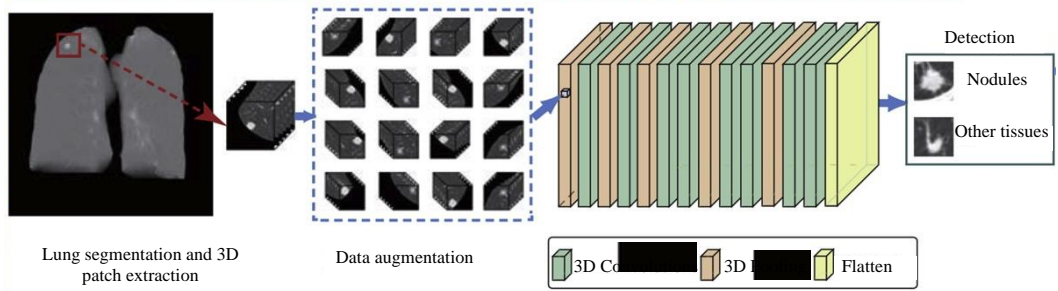

**Figure 7.** The basic procedure for lung nodule detection. The figure is modified from Reference [92].

To deal with the issue of limited data, Winkels and Cohen [93] proposed a 3D group convolutional neural network (3D-GCNNs). In this work, 3D rotations and reflections were used as input instead of translating a filter on the input (as in traditional 3D CNN). The authors showed that this approach needs only one-tenth of the data used in the conventional approach to obtain the same performance. In another work, Gong et al. [94] suggested a 3D CNN by exploiting the properties of ResNet and squeeze and excitation (SE) strategy. A 3D region proposal network using a UNet like structure was used for nodule detection, and then a 3D CNN was used for the reduction of false positives. The SE block increases the representation power of the network by focusing on channel-wise information. On the LIDC-IDRI dataset, 95.7% sensitivity was achieved with four false positives per scan. Pezeshk et al. [24] presented two-stage 3D CNN for automatic pulmonary nodule detection in CT scans. The first stage of 3D CNN was used for screening and candidate generation. The second stage was an ensemble of 3D CNNs trained with both positive and negative augmented patches.

The localization of biological architectures is a basic requirement for various initiatives in medical image investigation. Localization might be a hassle-free process for the radiologist, but it is usually a hard task for NNs that are vulnerable to variation in medical images induced by dissimilarities in the image acquisition process, structures and pathological differences among patients. Generally, a 3D volume is required for localization in medical images. Several techniques treat the 3D space as an arrangement of 2D orthogonal planes. Wolterink et al. [95] detected coronary artery calcium scoring in coronary CT angiography using a CNN based architecture. De Vos et al. [96] introduced the localization technique using a solitary CNN, and 2D CT image slices (chest CT, cardiac CT and abdomen CT) as inputs. While this work was related to a 3D localization approach, they did not use 3D CNN in a real sense. Further, the approach depended heavily on the accurate recognition of biological structures. Huo et al. [97] utilized the properties of a 3D fully connected CNN and presented a spatially localized atlas network tiles (SLANT) model for whole-brain segmentation on high-resolution multi-site images.

Intervertebral discs (IVDs) are modest joint parts that are located in between surrounding vertebrae and the localization of IVDs, which are usually important for spine disease analysis and measurement. In Reference [98], the authors presented a 3D detection for multiple brain structures in fetal neuro-sonography using fully connected CNNs and named it VP-Nets. They explained that the proposed strategy requires a comparatively less amount of data for training and can learn from coarsely annotated 3D data. Recently, a 3D CNN, based on regression, has been introduced in Reference [42] to assess the degree of enlarged perivascular spaces (EPVS) through 2000 basal ganglia scans from 3D head MRI. In Reference [99], the authors reported the human-level efficiency of 3D CNN in the landmark detection of clinical 3D CT data. In [100], Saleh et al. proposed a 3D CNN based regression models for 3D pose estimation of anatomy using T2 weighted imaging. They showed that the proposed network offers fine initialization for optimization-based techniques to increase the capture range of slice-to-volume registration. Xiaomeng et al. [101] presented fully connected, accurate and automatic 3D deep architecture for the localization and segmentation of IVDs using multimodal MR images. The work shows state-of-the-art performance in the MICCAI-2016 challenge for IVDs localization and segmentation section with a dice score of 0.912 for IVD segmentation. Cardiac magnetic resonance (CMR) imaging is popular in diagnosing various cardiovascular diseases. Vesel et al. [102] proposed a 3D DR-UNet (modified 3D UNet) for localization of cardiac structure in MRI volume. The model was evaluated on two datasets: the Automatic Cardiac Segmentation Challenge (ACDC) STACOM 2017, and Left Atrium Segmentation Challenge (LASC) STACOM 2018. Their model shows state-of-the-art results in terms of several performance indices.

*4.4. Registration*

Medical images of a single subject can be increasingly multi-modal in the same patient from CT, MRI T1 and MRI T2. Each imaging modality focuses on different features of the subject. Typically, a clinician is expected to view images of multiple modalities in different orientations to deduce a match between these images by visual comparison. The clinician is also expected to manually identify points in these images that have significant signal differences. Thus, two image analysis problems can be automated. First, the alignment or registration of datasets can be automated, and second, the automatic alignment of datasets can be made to modalities in which abnormalities are present. This allows us to identify prominent parts of an image for further review. In recent years, many efforts have been made in medical image registration using 3D deep learning. For example, Sokooti et al. successfully used 3D CNN for 3D nonrigid image registration in Reference [103]. For training the 3D CNN, 3D patches were extracted from 3D CT chest images. The network was trained on artificially generated displacement vector fields. The authors confirmed that their model outperformed a traditional B-spline registration method and performed on par with multi-resolution B-spline methods. However, for all the landmarks, multi-resolution B-spline methods outperformed their approach. Further, the capture range of their approach was limited to the size of the patches. A possible solution to increase the capture range of their method is to increase the size of the patches or to add more scales to the network. Torng et al. [104] showed the effectiveness of a shallow 3D CNN with three convolutional layers with filter sizes of 3x3, followed by 3D max-pooling layers and two fully connected layers with dropout for analyzing the interaction of amino acids to their neighboring microenvironment. The authors also proposed the CNN activation visualization technique called the atom importance map. For training and test data, 3D patches (local box) were extracted from the protein structure. Furthermore, the local structure was decomposed into five channels (inputs to 3D CNN), including oxygen, carbon, nitrogen and sulphur. The model shows a performance improvement of 20% when compared to the structure based on handcrafted biochemical features. To deal with the memory issue and computational cost, Blendowski and Heinrich [105] suggested a combination of MRF-based deformable registration and 3D CNN descriptors for lung motion estimation on non-rigidly deformed chest CT images.

There are several freely available medical image registration software and toolkits such as SimpleITK [106] and ANTs [107]. Typically, the registration process to these toolkits is done by iteratively updating the transformational parameters until a predefined similarity metric is optimized. These methods show a decent performance. However, their performance is limited by their slow registration process. To overcome this issue, several attempts have been made in the literature based on deep learning and 3D deep learning. Recently, Chee and Wu [108] used CNN as an affine image registration network (AIRNet) for MR brain image registration. AirNet, proposed by the authors, works in two parts, i.e., encoder and regressor. The architecture of the encoder part was drawn from DenseNet [109] with some modifications in filter structures (a mixture of 2D and 3D filters). The output of the encoder was then given to the regressor part. The proposed framework was compared to conventional registration algorithms used in the well-known software package SimpleITK [106]. The proposed framework shows significant improvements in $J_{ac}$ and $d_H$. The authors claim that this method was 100 times faster than other traditional methods. Zhou et al. [110] proposed 3D CNN for serial electron microscopy images (experiments were performed on two databases, Cremi and FIB25) registration. Recently, Zhao et al. [111] presented a 3D Volume Tweening Network (VTN) for 3D medical image (liver CT and brain MRI dataset) registration in an unsupervised manner. Compared to the traditional optimization approaches (ANTs [107], Elastix [112] and VoxelMorph-2 [113]), their method was 880 times faster, with state-of-the-art performance. In Reference[114], Wang et al. proposed a dynamic 2D/3D registration algorithm for accurate alignment between 2-D and 3-D images for fusion applications. The model introduced by the author was based on point-to-plane correspondence (PPC) and its dynamic registration procedure was fully capable of recovering 3-D motion from single-2D view images.

## 5. Challenges and Conclusions

It takes a large number of training samples to train deep learning models [53,115,116]. This is further strengthened by the recent successes of deep learning models trained on large datasets like the ImageNet. However, it is still ambiguous whether deep learning models can successfully work with smaller datasets, as in the case of medical images. The ambiguity is caused by the nature and characteristics of medical images. For example, the images from the ImageNet dataset possess large variations in their appearance (e.g., light, intensity, edges, color, etc.) [14,36,117–119] since the images were taken at different angles and distances, and have several different features that are completely different from medical images. Therefore, networks that need to learn meaningful representations of these images require large training parameters and thus training samples. However, in the case of medical images, there is much less variation in comparison to traditional image datasets [120]. In this regard, the process of fine-tuning of 3D CNN models, which are already trained on natural image datasets, can be applied to medical images [14,36,117–119,121,122]. This process, known as transfer learning, has been successfully applied to many areas of medical imaging.

Regardless of their high computational complexity, 3D deep networks have shown incredible performance in diverse domains. 3D deep networks require a large number of training parameters, especially in the case of 3D medical images, where the depth of the image volume varies from 20 to 400 slices per scan [36,79,123,124], with each scan containing very fine and important information about the patient. Usually, high-resolution scan volumes are of the size 512 × 512, and need to be downsampled before being fed into the 3D network to reduce the computational cost. Researchers generally use interpolation techniques to reduce the overall size of these medical image volumes, but come at the cost of significant information loss. There are also restrictions on the resizing of the medical image volume without the loss of significant information. This is still an unexplored area and there is further research scope.

While the number of trainable parameters of convolutional layers is independent of the input size, the number of trainable parameters in the subsequent fully connected layers depend on the output of the convolution layers. This often leads to intractable models due to a large number of trainable weights when the input images are fed into 3D CNN models without any down-sampling. However, this is not an issue in the case of 2D images, which have smaller latent representations that are learnt by convolution filters. This makes it harder (and more GPU intensive) to train 3D deep networks based on CNNs. The inception module by GoogLeNet can be further explored to address computational complexity in 3D medical image analysis. In recent times, many computational 3D imaging techniques have appeared in the literature where the acquired data is not necessarily a traditional image. Sometimes raw data may be suitable for a few applications in deep learning. For example, single-pixel imaging techniques are popular in unconventional applications, including X-rays. A brief review of various applications of single-pixel imaging in 3D reconstruction can be found in Reference [125]. Ghost imaging is also popular for image reconstruction, such as lens-less imaging and X-ray imaging. In order to enhance the quality of image reconstruction, Wang et al. [126] applied deep learning on the images reconstructed from traditional ghost imaging.

Indeed, in the deep learning context, learning the correct features might sound unconventional because we cannot be sure if the models learn features that are discriminating for the condition or just overfit on some specific features for the given dataset. CNNs can handle raw image data and they do not need to be handcrafted [11,117]. It is the responsibility of CNN to discover the right features from the data. While CNNs have made encoding the raw features in a latent space very convenient, it is important to understand whether the features learned by CNN are generalizable across datasets. Machine learning models often overfit on training samples as they only perform well on the test samples from the training dataset. This issue is acute in the case of medical imaging applications where there are issues with scanner variability, scan acquisition settings, subject demography, and heterogeneity in disease characteristics across subjects. Therefore, it is important to decode the trained network using model interpretability approaches and validate the important features learned by the network [127]. It also becomes important to report testing results with an external dataset whose samples were not used for training. However, this may not always be possible because of the paucity of datasets for training and testing.

Finally, the ultimate challenge is to go beyond a human-level performance. Researchers are working on reaching human-level performance for many tasks (known as Artificial General Intelligence) [35,53,128,129]. However, the lack of labeled images, high costs involved in labeling the datasets and lack of consensus among experts in validating the assigned labels [38,130,131] are some present challenges in the field. These issues force us to consider using reliable data augmentation methods and to generate samples with known ground-truths. In this regard, generative adversarial networks (GAN) [132], especially CycleGANs for cross-modal image synthesis, offer a viable approach for synthesizing data. They are being used to produce pseudo images that are highly similar to the original dataset.

**Author Contributions:** All authors conceptualized the ideas and conducted the literature search, prepared the figures, tables, and drafted the manuscript. All authors have read and approved the manuscript.

**Funding:** Authors acknowledge the support from Lee Kong Chian School of Medicine and Data Science and AI Research (DSAIR) center of Nanyang Technological University Singapore (Project Number ADH-11/2017-DSAIR).

PP and BG also acknowledges the support from the Cognitive Neuro Imaging Centre (CONIC) at Nanyang Technological University Singapore.

**Conflicts of Interest:** The authors declare no conflict of interest.